# A Multi-Agent Approach to Optimal Sizing of a Combined Heating and Power Microgrid

Soheil Mohseni
Department of Electrical Engineering, University Campus 2,
University of Guilan
Rasht, Iran
soheilmohseni92@gmail.com

Seyed Masoud Moghaddas Tafreshi
Department of Electrical Engineering, Faculty of
Engineering, University of Guilan
Rasht, Iran
tafreshi@guilan.ac.ir

*Abstract*—In this paper, a novel multi-agent based method is applied to the problem of optimal sizing of the components of an islanded combined heating and power residential microgrid such that the residential power and heat demands and hydrogen refilling demands of the fuel cell electric vehicles (FCEVs) are met. It is assumed that the proposed multi agent-based architecture consists of five different agents, namely generation agent (GA), electrical and thermal loads agent (LA), FCEV refilling station agent (SA), control agent (CA), and design agent (DA) which are organized in three levels. Design agent (DA) is the main agent of the proposed multi-agent system (MAS) that according to its interactions with CA and by minimizing the total costs of the system through particle swarm optimization (PSO), finds the optimal sizes of the system's components. This study is performed for a typical stand-alone combined heating and power microgrid.

## I. Introduction

Sustainable energy systems have received global attention in many countries according to the depletion of fossil fuels, global warming and climate change issues. Renewable energy sources (RESs) such as wind and solar are well accepted sustainable energy sources [1]. A microgrid is a low-voltage distribution system with distributed energy resources, storage devices, and controllable loads that are operated either islanded or connected to the main power grid in a controlled way. The steady progress in the development of distributed power generation, such as microgrids and renewable energy technologies, are opening up new opportunities for the utilization of distributed energy resources [2].

Optimal sizing of hybrid micro-combined heat and power (CHP) systems defined on the basis of linear programming techniques have been addressed in [3] that takes advantage of rapid calculations even in the presence of a high number of variables. The optimal design of microgrids with CHP generators, through the development of a mixed-integer linear programming (MILP) model has been addressed in [4] that has considered both environmental and economic concerns. A MILP-based optimal structural design model of residential power and heat supply devices has been developed in [5] that considers the operational and capital recovery constraints of these devices. A decision model for the planning of the energy production has been proposed in [6] that aims to find the optimal sizes of the components of a smart district that satisfy its electrical and thermal demands.

In this paper, the optimal sizing problem of a combined heating and power residential microgrid is considered which is equipped with photovoltaic/wind turbine/fuel cell generation, electrolyzer and hydrogen storage, boiler and heater, and a FCEV refilling station. For this purpose, a multi-agent system is proposed that finds the optimal sizes of the components by mean of particle swarm optimization (PSO). The considered energy management strategy in the proposed MAS controls the refilling of FCEVs depending on the hour of the day to reduce cost and avoid overload during peak hours.

The rest of this paper is organized as follows: In section II, configuration of the microgrid is presented. Section III describes the multi-agent architecture used for optimal sizing along with the description of components belonging to each agent. Section IV presents the simulation results of the MAS-based architecture used for optimal sizing of the microgrid. Finally, the conclusion of this study is presented in section V.

## II. Configuration of the Microgrid

Power flow diagram of the proposed microgrid is shown in Fig. 1. According to this diagram, the output power of the photovoltaic unit is equal to $P_{PV}$ that depends on the solar radiation and the output power of the wind generation unit is equal to $P_{wg}$ that depends on the wind speed. When the summation of the output powers of PV arrays and wind turbines is more than electricity demand ($P_{load}$), the first priority of using this surplus power ($P_{ren-el}$) is producing hydrogen in the electrolyzer and storing the hydrogen in the tank. If the surplus power exceeds electrolyzer's rated power or the tank becomes full, the surplus power ($P_{ren-h}$) will be used for heat production in the heater to supply the thermal load to the amount of $Q_{h-tl}$. In this case, if there is shortage of heat production, boiler compensates it. On the other hand, when the summation of the output powers of the PV arrays and wind turbines is less than the electricity demand, fuel cell consumes the hydrogen stored in the tank to the amount of $P_{tank-fc}$ and compensates the shortage of the power production ($P_{fc-conv}$). If the shortage of power exceeds fuel cell's rated power or the stored hydrogen cannot afford that, some fraction of the electricity demand of the microgrid must be shed which leads to loss of load. In this case, if the heat production of the fuel cell does not satisfy the thermal load ($Q_{load}$), boiler will compensate the shortage of the heat production. Finally, if the power production of renewable sources ($P_{PV}+P_{wg}$) is equal to the electrical load demand, all of the renewable generations will be delivered to the electrical loads and boiler should satisfy the thermal loads. In all of these cases, FCEVs can be refilled at the station and consume the hydrogen power of $P_{tank-sta}$.

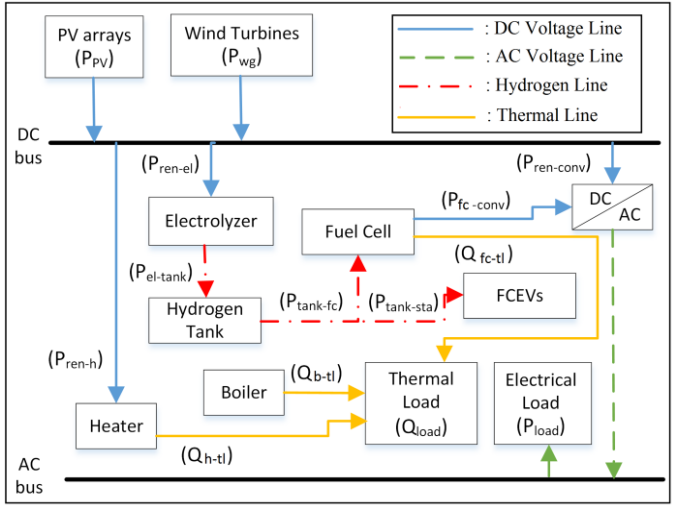

Fig. 1. Schematic diagram of the proposed microgrid system

### III. MULTI-AGENT SYSTEM

The considered MAS for optimal sizing of the proposed microgrid has five agents, namely generation agent, electrical and thermal loads agent, refilling station agent, control agent, and design agent which are organized in three levels. All the agents associated with the generation or consumption of electricity, heat, and hydrogen belong to the field level, which is the lowest level of the considered architecture for the MAS. GA, LA, and SA belong to this level. In the coordination level, coordination of the generation of electricity, heat, and hydrogen with consumption of the electrical, thermal, and hydrogen loads obtains. CA belongs to this level. Design level is the highest level of the proposed architecture that finds the optimal sizes of the components according to the interactions with coordination level such that the residential power and heat demands, and FCEVs' hydrogen refilling demands are met. In the proposed MAS, DA belongs to this level.

#### A. Generation agent

This agent is responsible for management of the resources of the microgrid. Generation resources of the microgrid include PV arrays, wind turbines, and the fuel cell. Also, this agent has an electrolyzer to provide hydrogen for the fuel cell and uses a hydrogen tank as a storage unit.

*1) Components under the supervision of generation agent:* Modeling of the components which are under the supervision of generation agent is presented in the following subsections.

*a) PV arrays:* The output power of the PV generator can be calculated according to the following equation [7]:

$$P_{PV} = \eta_g N_{PV} A_m G_t, \quad (1)$$

where $\eta_g$ is the instantaneous PV generator efficiency that is considered to be 15.4% in this paper [7], $A_m$ is the area of a single module used in the system (m$^2$) which is considered to be 1.9 m$^2$ in this paper [7], $G_t$ is the total global irradiance incident on the titled plane (W/m$^2$), and $N_{PV}$ is the number of modules. In this analysis, each PV array has a rated power of 1 kW. The capital cost of 1 unit is $2000 [7], while the replacement and maintenance costs are taken as $1800, and $0/yr, respectively. The lifetime of a PV array is taken to be 20 years [8].

*b) Wind turbine:* The power of the wind turbine ($P_{wg}$ (kW/m$^2$)) can be described in terms of wind speed by (2) [9].

$$\begin{cases} 0 & V < V_{cin}, V > V_{coff} \\ P_{wg-max} \times ((V - V_{cin})/(V_r - V_{cin}))^3 & V_{cin} \leq V < V_r \\ P_{wg-max} \times \dfrac{P_{furl} - P_r}{V_{coff} - V_r} \times (V - V_r) & V_r \leq V \leq V_{coff} \end{cases}, \quad (2)$$

where $V_{cin}$ is the cut in wind speed (m/s) that is considered to be 2.5 m/s, $V_{coff}$ is the cut off wind speed (m/s) that is considered to be 25 m/s, $V$ is the wind speed (m/s), $V_r$ is the nominal wind speed (m/s) that is considered to be 11 m/s, $P_{rated}$ is the nominal power of wind turbine which is considered to be 1 kW, $P_{wg-max}$ is the maximum power of wind turbine, $P_{wg-min}$ is the minimum power of wind turbine, and $P_{furl}$ is the power of wind turbine in cut off wind speed. Actual power available from wind turbines is given by (3).

$$P_{wg}(t) = \eta_w N_{wg} A_w P_w(t), \quad (3)$$

where $\eta_w$ is the efficiency of the wind turbine generator and corresponding converters, $N_{wg}$ is the number of wind turbine generators, and $A_w$ is the swept area. The capital cost of 1 unit is $1500 [7], while the replacement and maintenance costs are taken as $900, and $45/yr, respectively. The lifetime of a wind turbine is taken to be 20 years [8].

*c) Fuel cell:* Proton exchange membrane fuel cell is an environmentally clean power generator which combines hydrogen fuel with oxygen from air to produce heat and electricity. The efficiency of the fuel cell is fed to the computational program as the input. Fuel cell's power and heat outputs can be defined by the following equations:

$$P_{fc-conv} = P_{tank-fc} \times \eta_{fc,el}, \quad (4)$$

$$Q_{fc-tl} = P_{tank-fc} \times \eta_{fc,th}. \quad (5)$$

where $\eta_{fc,el}$ and $\eta_{fc,th}$ are the electeical and thermal efficiencies of the fuel cell and considered to be 40% and 50%, respectively. The capital, replacement, and maintenance costs are taken as $2000 [7], $1500, and $100/yr for a 1 kW system, respectively. The FC's lifetime is considered to be 5 years [8].

*d) Electrolyzer*: Electrolysis to dissociate water into its separate hydrogen and oxygen constituents has been in use for decades, primarily to meet industrial chemical needs.

The electrolyzer's output power can be calculated by the following equation:

$$P_{el\text{-}tank} = P_{ren\text{-}el} \times \eta_{el}, \qquad (6)$$

In this analysis, a 1 kW system is associated with $1500 capital [7], $1000 replacement, and $15/yr maintenance cost. The electrolyzer's lifetime and efficiency are considered to be 20 years and 75%, respectively [8].

*e) Hydrogen tank:* The energy of hydrogen stored in the tank at time step *t* is obtained by the following equation:

$$E_{tank}(t) = E_{tank}(t\text{-}1) + P_{el\text{-}tank}(t) \times \Delta t - (P_{tank\text{-}fc}(t) + P_{tank\text{-}sta}(t)) \times \eta_{storage} \times \Delta t, \qquad (7)$$

where $\eta_{storage}$ is the efficiency of the storage system which is assumed to be 95%, and $\Delta t$ is the duration of each time step which is equal to one hour in this study.

To calculate the mass of the stored hydrogen in the tank, the following equation can be used [8]:

$$m_{storage}(t) = E_{tank}(t) / HHV_{H2}, \qquad (8)$$

where $HHV_{H2}$ is the higher heating value of hydrogen which is equal to 39.7 kWh/kg [8]. It is worth mentioning that there are lower and upper limits for amount of the stored hydrogen. It is not possible that the mass of stored hydrogen exceeds the rated capacity of the tank. On the other hand, because of some problems, e.g. hydrogen pressure drop, a small fraction of the hydrogen (here, 5%) may not be extracted. This fraction is the lower limit of the stored hydrogen.

In this analysis, a 1 kg system is associated with $500 capital [7], $450 replacement, and $5/yr maintenance cost. The hydrogen tank's lifetime and efficiency are considered to be 20 years and 95%, respectively [8].

*f) Heater:* The equipments that have heating elements are called heaters. In this analysis, heater is used for supplying the heat load of the microgrid when there is surplus power generation. Heater's output power can be defined by the following equation:

$$Q_{h\text{-}tl} = P_{ren\text{-}h} \times \eta_h. \qquad (9)$$

where $\eta_h$ is the efficiency of the heater. In this analysis, a 1 kW system is associated with $281 capital [10], $150 replacement, and $5/yr maintenance cost. The heater's lifetime and efficiency are considered to be 20 years and 90%, respectively [10].

*g) Boiler:* It is assumed that boiler's fuel is natural gas. Equation (10) shows the heat output of the boiler based on input fuel and (11) shows the cost of the input fuel per kWh.

$$Q_{b\text{-}tl} = C_{in} \times \eta_b, \qquad (10)$$

$$C_{boiler} = C_{in} \times C_{fuel}. \qquad (11)$$

where $\eta_b$ is the boiler's efficiency, $C_{in}$ is the input fuel of the boiler and $C_{fuel}$ is the fuel cost which is equal to $0.03/kWh in this study. In this analysis, a 1 kW gas boiler is associated with $85 capital [11], $60 replacement, and $2/yr maintenance cost. The boiler's lifetime and efficiency are considered to be 15 years and 94%, respectively [11].

*h) DC/AC converter:* The power electronic circuit used to convert DC into AC is known as inverter. In this analysis, a 1 kW system is associated with $700 capital [7], $650 replacement, and $7/yr maintenance cost. The converter's lifetime and efficiency are considered to be 15 years and 90%, respectively [8].

*2) Simulation algorithm of the generation agent:* This agent, by coordination of the control agent, interacts with LA and SA in order to supply the electricity, thermal, and hydrogen demands. In this regard, first CA sends the sizes of PV arrays, wind turbines, electrolyzer, hydrogen tank, fuel cell, boiler, heater, and DC/AC converter that are determined by the DA, to the GA and at each time step *t* asks it for the next hour value of the output power of solar and wind generations. Then, GA sends the requested data to the CA. In this paper, it is assumed that GA acts such that when it receives the request of storing the surplus power from the CA, it will use the excess energy in electrolyzer and stores the produced hydrogen in the tank. On the other hand, when the GA receives the request of supplying the shortage of electrical power production from the CA, it uses the stored hydrogen in the fuel cell to compensate the shortage of the power production. Also, when it receives the request of supplying the shortage of thermal load, it turns the boiler on.

## B. Electrical and Thermal loads agent

Electrical and thermal loads agent is a simple reflex agent that aggregates all the electricity and heat consumptions of the residential loads. The electrical loads that LA is responsible for forecasting them are interruptible ($P_i$) and uninterruptible ($P_{uni}$) loads and should be supplied subject to a reliability constraint. It is assumed that at each hour, 15% of loads are interruptible and 85% of them are uninterruptible.

## C. FCEV refilling station agent

Refilling station agent is responsible for refilling the FCEVs that arrive at the station and sends the forecasted refilling demand to the control agent. In this paper, hydrogen pipeline transport is used to deliver hydrogen from the hydrogen tank of the GA to the station. The next stage is to compress the hydrogen from 1 bar to 700 bar and pump the gas to storage vessels for delivery to the fuel pump. FCEVs are designed to accept 350 or 700 bar with the greater pressure adding to the cost per kg. Therefore, a hydrogen compressor is needed at the station. Also, in order to fuel 700 bar fuel cell vehicles, a hydrogen dispenser is needed.

In this study, Pure Energy Centre's hydrogen compressor and dispenser are used to achieve the 700 bar hydrogen needed for refilling the FCEVs. This compressor has the maximum power of 200 kW which is enough for supplying the FCEVs that arrive at the station. It is assumed that the size of hydrogen compressor-dispenser system is fixed and does not participate in the optimal sizing problem. The capital cost of this system is $100,000 [12], while the replacement and maintenance costs are taken as $80,000 and $200/yr, respectively. The lifetime and efficiency of this system are considered to be 20 years and 49%, respectively [13].

*1) Modeling of FCEVs:* In this paper, Toyota Mirai is chosen as a mid-size sedan FCEV which is manufactured by Toyota. Mirai is equipped with a 113 kW fuel cell-powered electric motor and two hydrogen tanks with a three-layer structure made of carbon fiber-reinforced plastic consisting of nylon 6 from Ube Industries, and other materials. The tanks store hydrogen at 700 bar and have a combined weight of 87.5 kg and 5 kg capacity. The Mirai refueling takes between 3 and 5 minutes and Toyota has expected a total range of 300 miles on a full tank.

*2) Simulation algorithm of the station agent:* It is assumed that unmanaged refilling profile of the station is available for a year and FCEVs can be managed to be refilled in the evening and night-time off-peak hours, thereby the total cost of the microgrid reduces. This method is such that at each step time $t$, the SA sends the hydrogen demand to the CA and when it receives the request of the reduction of the demand from the CA, it calculates the deferrable hydrogen loads and adds them to the next hour demand and then sends the updated demand of the $t$th hour to the CA. Also, if it cannot refill some of the FCEVs by the allocated hydrogen power of the CA, it sends the information of unsupplied hydrogen demand to the CA.

### D. Control agent

Control agent coordinates the interactions between the generation agent and electrical and thermal loads agent and station agent. This agent, after receiving a request for operating the microgrid with determined sizes from DA and sending the related sizes to the GA and SA, at each time step $t$, requests GA for the next hour solar and wind power productions, requests LA for the next hour electricity and thermal demands, and requests SA for the next hour hydrogen demand needed for refilling the FCEVs. After receiving the information on the operation of the generation and storage components of the microgrid from the GA, it establishes the energy management strategy that is presented in section II and when it finds that some of the electrical loads should be interrupted for balancing the generation and demand, it interrupts them and when it finds that some of the hydrogen loads can be deferred to the evening and night-time off-peak hours, it manages them to be refilled at those hours. Finally, CA sends the information on the operation of the microgrid to the DA.

### E. Design agent

Design agent is responsible for optimal sizing of the components of the microgrid.

*1) System's cost:* In this paper the capital, replacement, and maintenance costs of each component of the microgrid have been considered and net present cost (NPC) is chosen for the calculation of the system's cost. The NPC for a specific device can be expressed by the following equation:

$$NPC_i = N_i \times (CC_i + RC_i \times K_i + MC_i \times PWA(ir,R)), \quad (12)$$

where:

| | |
|---|---|
| $N$ | Number of units and/or the unit capacity (kW or kg) |
| $CC$ | Capital investment cost ($/unit) |
| $K$ | Single payment present worth |
| $RC$ | Replacement cost ($/unit) |
| $MC$ | Maintenance and repair cost ($/unit.yr) |
| $PWA$ | Present worth annual payment |
| $ir$ | Real interest rate (6%) |
| $R$ | Project lifetime (20 years) |

*2) The objective function:* The objective function is sum of all net present costs:

$$NPC = NPC_{PV} + NPC_{WT} + NPC_{el} + NPC_{tank} + NPC_{FC} + \\ NPC_{boiler} + NPC_{heater} + NPC_{conv} + NPC_{sta} + NPC_{em} + \\ NPC_{fuel} + NPC_{Pi} + NPC_{Puni} + NPC_Q + NPC_h, \quad (13)$$

where $NPC_{Pi}$, $NPC_{Puni}$, $NPC_Q$, and $NPC_h$ are the $NPC$s of interrupting the interruptible electrical loads, uninterruptible electrical loads, thermal loads, and hydrogen loads, respectively. It is worth mentioning that the penalty factors for unsupplied uninterruptible electrical loads and thermal loads are considered to be 5.6 $/kWh and the penalty factors for unsupplied interruptible electrical loads and hydrogen loads are considered to be 0.56 $/kWh.

*3) Reliability:* In order to evaluate the reliability of supplying the electrical and thermal loads of the microgrid, the equivalent loss factor (ELF) index is used and these indices can be calculated by the following equations:

$$ELF_{el} = (1/8760) \sum_{t=1}^{8760} (P_{Uns\_uni}(t) / P_{uni}(t)), \quad (14)$$

$$ELF_{th} = (1/8760) \sum_{t=1}^{8760} (Q_{Uns}(t) / Q(t)), \quad (15)$$

where $P_{Uns\_uni}(t)$ and $Q_{Uns}(t)$ indicate the unsupplied uninterruptible electrical loads and unsupplied thermal loads at time step $t$, respectively.

*4) Emission cost:* Emission cost function can be calculated by the following equation [14]:

$$NPC_{em} = PWA(ir,R) \times \sum_{t=1}^{8760}\sum_{i=1}^{N}\sum_{k=1}^{M} \alpha_k (EF_{ik} P_i), \quad (16)$$

where $\alpha_k$ is the externality costs of emission type $k$, $EF_{ik}$ is the emission factor of generation unit $i$ and emission type $k$, $N$ is the number of generation units and $M$ is the emission type ($NO_x$, $CO_2$ or $SO_2$). Externality costs and emission factors of the boiler are shown in Table I [14].

TABLE I. EXTERNALITY COSTS AND EMISSION FACTORS

| Emission type | Externality costs ($/lb) | Emission factors of the boiler (lb/MWh) |
|---|---|---|
| $NO_X$ | 4.2 | 5.06 |
| $SO_2$ | 0.99 | 11.9 |
| $CO_2$ | 0.014 | 1965 |

*5) fuel cost:* Net present cost of the input fuel of the boiler can be calculated according to the following equation:

$$NPC_{fuel} = PWA(ir,R) \times \sum_{t=1}^{8760} \frac{Q_{boiler}(t)}{\eta_{boiler}} \times C_{fuel}. \quad (17)$$

*6) Simulation algorithm of the design agent:* After the determination of the sizes of the components through PSO by this agent, the determined sizes of the components will be sent to the CA. Accordingly, CA sends the determined sizes of the components to the GA. Then, CA coordinates the GA with LA and SA and sends the information on the operation of the microgrid to the DA. After receiving this information, DA calculates the objective function and ELF indices. If $ELF_{el}$ and $ELF_{th}$ are less than 0.01, and the energy content of the tank at the end of a year is not less than its initial amount, then DA selects the obtained sizes as optimum sizes.

IV. SIMULATION RESULTS

In this section, the proposed MAS-based architecture for optimal sizing of the components of the proposed islanded combined heating and power microgrid is simulated and the optimum combination of the components is calculated. The simulation of the proposed MAS is performed using MATLAB software.

The annual solar radiation and wind speed data that have been considered in the simulation program are shown in Fig. 2 and Fig. 3, respectively. It is obvious that sun irradiance is equal to zero at nights but because of the scale of the figure, it cannot be seen in Fig. 2.

In obtaining the hydrogen load curve of the FCEV station, it is assumed that there are 150 FCEVs in the microgrid that should be refilled two times in a week.

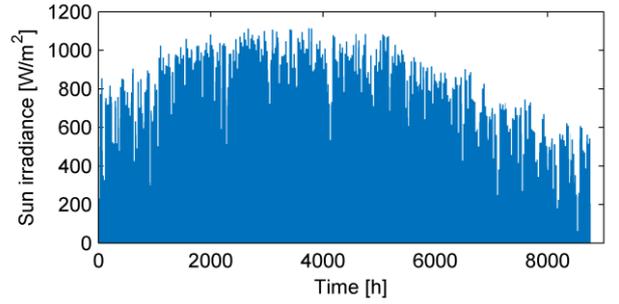
Fig. 2. Annual sun irradiance

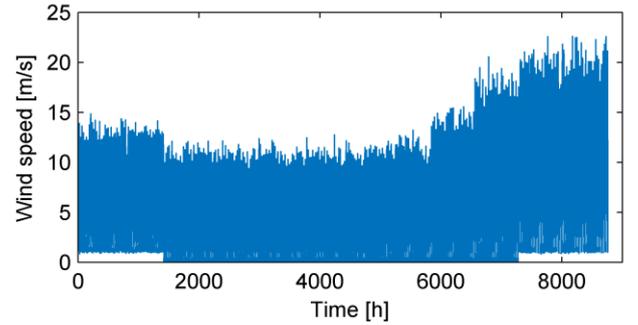
Fig. 3. Annual wind speed

The considered annual electrical and thermal load curves are shown in Fig. 4 and Fig. 5, respectively.

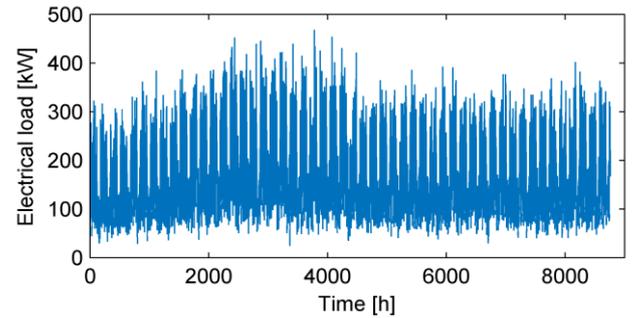
Fig. 4. Annual electrical load curve of the microgrid

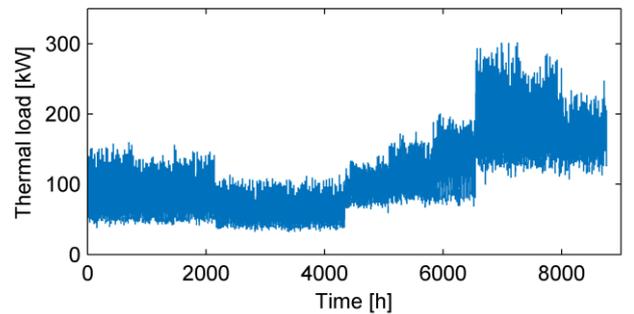
Fig. 5. Annual thermal load curve of the microgrid

Simulations are done in two scenarios. In scenario1, electrical and hydrogen demands of the microgrid are considered as fixed loads and the MAS is simulated without controlling them. In scenario2, it is assumed that the considered energy management strategy in the proposed MAS controls the supplying of the interruptible loads and refilling of the FCEVs depending on the hour of the day.

*A. Scenario1:* In this scenario, hydrogen refilling demand is considered as a fixed load and all of the electric loads are uninterruptible loads. Results of optimal sizing problem through the MAS in this case are shown in Table II. In this scenario, the total cost is equal to $36,547,871.

TABLE II. OPTIMAL SIZES OF THE COMPONENTS IN SCENARIO1

| PV | WT | Electrolyzer (kW) | Tank (kg) |
|---|---|---|---|
| 2851 | 232 | 1391 | 1563.03 |
| Fuel cell (kW) | DC/AC (kW) | Boiler (kW) | Heater (kW) |
| 389.07 | 391.43 | 254.32 | 31.87 |

*B. Scenario2:* In this scenario, hydrogen refilling demand is considered as a deferrable load and at each hour, 15% of the electrical loads are considered as interruptible loads. Results of optimal sizing problem through the MAS in this case are shown in Table III. In this scenario, the total cost is equal to $34,911,246.

TABLE III. OPTIMAL SIZES OF THE COMPONENTS IN SCENARIO2

| PV | WT | Electrolyzer (kW) | Tank (kg) |
|---|---|---|---|
| 2593 | 308 | 1498 | 1529.74 |
| Fuel cell (kW) | DC/AC (kW) | Boiler (kW) | Heater (kW) |
| 312.55 | 329.76 | 98.30 | 179.22 |

Results show that using the proposed deferrable refilling method and benefitting from the potential of interruptible loads in the MAS, decreases the optimal size of PVs, fuel cell, converter, and boiler and increases the optimal size of the WTs, electrolyzer, and heater, which in turn decreases the total cost of the system. This is because of that the summation of the costs of reduced number of PVs, decreased capacity of fuel cell, converter, and boiler, and decreased amounts of pollution penalties is more than the summation of the costs of added WTs and higher capacity electrolyzer and heater. In the proposed MAS, the reason behind decreasing the number of PVs is that it uses a control strategy that avoids overload during peak hours, the reason behind increasing the number of WTs is that the majority of the FCEVs are going to be charged in the evening and night-time off-peak hours when solar energy is not available. Furthermore, the reason behind increasing the capacity of the heater and decreasing the capacity of the boiler is that by interrupting some of the electrical loads and deferring the refilling of the FCEVs, it would be possible to use the surplus power of the PVs and WTs for supplying the thermal loads by heater and therefore, the amounts of pollution penalties that the boiler is responsible for them decrease significantly. The reason of the decreased capacity of the fuel cell and converter is that in the second scenario, it is assumed that 15% of the electrical loads are interruptible at each hour. Moreover, the capacity of the hydrogen tank does not change very much and the proposed energy management strategy has not affected its size.

## V. CONCLUSION

In this paper, a multi-agent system for optimal sizing of an islanded combined heating and power microgrid has been developed. It is observed that according to the considered energy management strategy in the proposed MAS and interrupting some of the electrical loads and deferring the refilling of the FCEVs' tanks, overloading can be avoided, available components can be utilized better and therefore, a lower NPC for a 20 year investment horizon achieves.